\documentclass[reprint,
superscriptaddress,
amsmath,amssymb,
aps,
pra,
]{revtex4-2}

\usepackage{color}
\usepackage{graphicx}
\usepackage{dcolumn}
\usepackage{bm}
\usepackage{diagbox}
\usepackage{listings}
\usepackage{xcolor}
\usepackage{hyperref}
\hypersetup{
    colorlinks=true,
    linkcolor=cyan,           
    citecolor=cyan,           
    urlcolor=cyan
}

\usepackage[normalem]{ulem}

\usepackage{fncylab}
\labelformat{equation}{(#1)}
\labelformat{section}{(#1)}

\usepackage[frozencache,cachedir=.]{minted}

\usepackage[format=plain,labelfont={bf,it}]{caption}
\newenvironment{code}{\captionsetup{type=listing}}{}
\renewcommand\listingscaption{Codeblock}
\usepackage[font=small,labelfont=bf,justification=justified]{caption}
\setminted[python]{ bgcolor=gray!6, fontfamily=tt, gobble=0, fontsize=\footnotesize,
breaklines=true,
    framerule=0.4pt,
    funcnamehighlighting=true, tabsize=2, obeytabs=false, mathescape=false samepage=true,
    showspaces=false, showtabs =false, texcl=false, linenos=false, xleftmargin=\parindent,
    numbersep=1pt, texcomments }

\newmintedfile[pythonfile]{python}{}

\DeclareUnicodeCharacter{2212}{-}

\begin{document}
\preprint{APS/123-QED}

\title{OpenQAOA --- An SDK for QAOA}

\author{Vishal Sharma}
\author{Nur Shahidee Bin Saharan}
\author{Shao-Hen Chiew}
\author{Ezequiel Ignacio Rodríguez Chiacchio}
\author{Leonardo Disilvestro}
\author{Tommaso Federico Demarie}
\author{Ewan Munro}
\affiliation{Entropica Labs, 186b Telok Ayer Street, Singapore 068632}
\email{openqaoa@entropicalabs.com}

\begin{abstract}
We introduce OpenQAOA, a Python open-source multi-backend Software Development Kit to create, customise, and execute the Quantum Approximate Optimisation Algorithm (QAOA) on Noisy Intermediate-Scale Quantum (NISQ) devices and simulators. OpenQAOA facilitates the creation of QAOA workflows, removing the more tedious and repetitive aspects of implementing variational quantum algorithms. It standardises and automates tasks such as circuit creation across different backends, ansatz parametrisation, the optimisation loop, the formatting of results, and extensions of QAOA such as Recursive QAOA. OpenQAOA is designed to simplify and enhance research on QAOA, providing a robust and consistent framework for experimentation with, and deployment of, the algorithm and its variations. Importantly, a heavy emphasis is placed on the provision of tools to enable QAOA computations at the scale of hundreds or thousands of qubits.
\end{abstract}

\maketitle 

\section{QAOA with hundreds of qubits}

Variational quantum algorithms (VQAs) are a firmly established hybrid approach to using noisy intermediate scale quantum (NISQ) and classical computers in tandem to tackle problems of scientific and industrial interest~\cite{mcclean2016theory,cerezo2021variational, Bharti_2022}. In contrast to quantum algorithms such as Grover’s or Shor’s algorithms, which come with performance guarantees and even speed-ups over classical approaches, VQAs are heuristic methods whose performance has become a major line of investigation in recent years. Today’s quantum computers typically support only a few tens of qubits. However, this number is expected to rise to several hundred in the next few years. Hence, research and development with VQAs will soon enter a new phase, characterised by a marked shift in scale. \newline

This upwards step in scale will likely precipitate significant changes in research, working methods, and priorities within the field. For example, classical simulations of VQAs currently constitute a large part of the algorithmic prototyping and experimentation process. Moving forward, a more significant proportion of this work will take place directly on quantum computers. Meanwhile, on the applications side, the focus will shift away from small and illustrative ‘toy’ problems towards ones that capture more of the complexity of valuable target use-cases.\newline

One of the most prominent VQAs is the Quantum Approximate Optimisation Algorithm (QAOA)~\cite{Farhi2014}, which has attracted much attention for its potential application to combinatorial optimisation problems. A broad spectrum of research work on QAOA has emerged, attempting to clarify questions ranging from fundamental aspects of its mechanism and its limitations~\cite{Hastings_QAOA, farhi_qaoa_whole_graph, Zhou2020, reachability_deficits, Bravyi2020, mcclean2021low, stilck2021limitations, symmetries_QAOA} to its suitability for specific use-cases~\cite{vikstaal2020applying,stollenwerk2020toward,dalyac2021qualifying,QAOA_knapsack, zhang2021qed, bentley2022quantum,brandhofer2022benchmarking, Quera_MIS}. Coupled with the imminent arrival of quantum computers with hundreds of qubits, the growth in research activity for QAOA motivates the need for robust, consistent supporting software tools.\newline

\begin{figure}
\includegraphics[width=\textwidth/2]{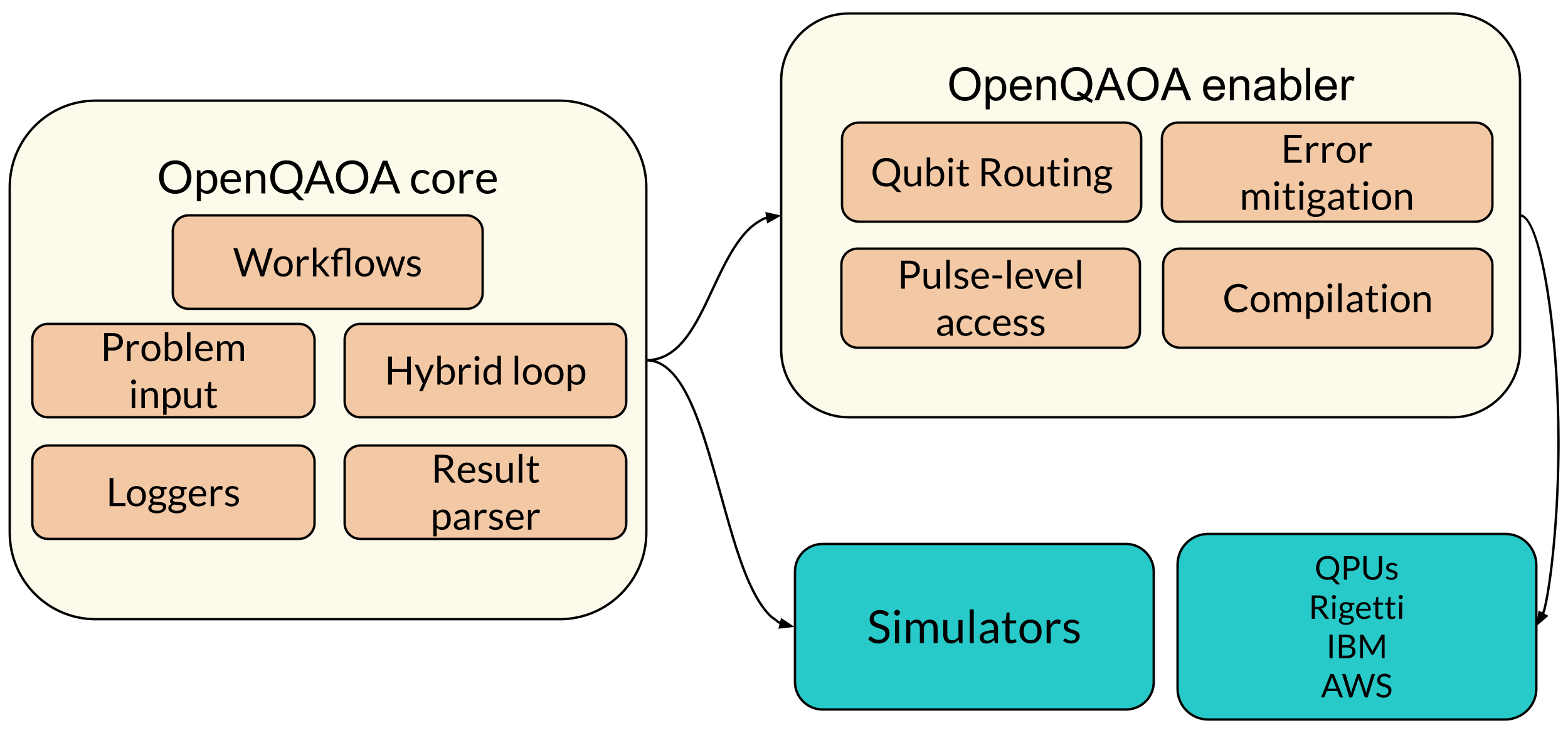}
\caption{OpenQAOA is conceptually divided into two parts. The Core contains all the features needed to set up and manage a QAOA computation, from problem input to result retrieval. The Enabler contains tools to support workflow deployment and execution on quantum hardware, with the aim of providing the features necessary to execute QAOA at the scale of hundreds of qubits.}
\label{fig:qaoa_structure}
\end{figure}


In this article, we introduce OpenQAOA, an open-source Python Software Development Kit (SDK) that aims to meet these needs by simplifying and enhancing research work for all user groups. OpenQAOA has a strong focus on integrating physical devices, providing support for multiple quantum computing backends, as well as simulators. Algorithmically, OpenQAOA provides robust implementations of multiple extensions to the original proposal of Farhi et al.~\cite{Farhi2014}, including alternative circuit parametrisations and mixer operators, as well as strategies such as Recursive QAOA~\cite{Bravyi2020}. \newline

OpenQAOA is more than a library to solve optimisation problems with QAOA: it strives to be a tool to bridge algorithms and hardware. Executing computations on real quantum hardware at scale introduces many challenges, such as the need for effective circuit compilation methods to minimise the impact of noise. OpenQAOA seeks to address these issues through custom QAOA-tailored solutions, as well as integration with existing tools in the quantum open source ecosystem.\newline


The source code for OpenQAOA is licensed under Apache 2.0 and is available at \url{https://github.com/entropicalabs/openqaoa}. The latest documentation can be consulted at \url{https://el-openqaoa.readthedocs.io/en/latest/}. 

\section{An SDK for QAOA at scale}

The role of an SDK is to provide a set of software tools and programs to create specific applications, resulting in a faster development environment for building and designing reliable and performant code. Within the context of quantum optimisation in the NISQ era, this means abstracting repeated and tedious tasks, improving cross-platform development, and providing tools and features that enable computations at the scale of hundreds or thousands of qubits. To illustrate different types of obstacles on the path to large-scale QAOA, and how OpenQAOA seeks to address them, we give three specific examples. \newline

Firstly, a QAOA circuit can rarely be executed directly, `as-is' on a quantum processing unit (QPU). The abstract circuit must be compiled to conform to the specification of the target device, taking into account its connectivity and available native gates. Furthermore, this compilation should minimise metrics strongly associated with the proliferation of noise, such as the circuit depth and the number of two-qubit gates. This \emph{noise suppression} challenge will become more acute as researchers focus on increasingly larger problems. Therefore, in the NISQ era, circuit construction must be `hardware-aware', and should call on suitable optimisation methods to limit the impact of noise. A major goal of OpenQAOA is to achieve tight integration with powerful circuit compilation tools. \newline

Secondly, a greater degree of algorithmic flexibility will be needed to tackle larger and more complex problems. QAOA has many customisable components, including the circuit initial state, the mixer operator, the ansatz parametrisation, the initial parameter values, and the classical optimiser. The ability to easily experiment with these components, individually and in combination, will facilitate research on a wide range of problem classes and provide extensive scope for optimising performance at scale. OpenQAOA aims to meet the need for algorithmic design flexibility through its highly modular structure.  \newline

Finally, the performance of QAOA may be improved for large problems by embedding the algorithm in a wider workflow. A prominent example is Recursive QAOA (RQAOA), which combines QAOA with a divide-and-conquer approach, whereby a series of successively smaller problems are defined by eliminating variables \cite{Bravyi2020}. Eliminating variables (qubits) one by one -- as contemplated in the original RQAOA proposal -- may lead to impractically long runtimes, especially for problems initially defined on many hundreds or thousands of variables. Besides a standard RQAOA workflow, OpenQAOA provides a customisable adaptive strategy, Ada-RQAOA, allowing multiple qubits to be eliminated at a time \cite{EzeInProgressPaper}. \newline

Solving these challenges is an effort to lower the entry barrier to quantum computing, while maintaining a research-grade code base fit for executing QAOA at scale. The role of an SDK such as OpenQAOA can then be seen as a democratising force, bringing quantum applications closer to researchers and end users.

\section{QAOA -- The basics}
\label{sec:QAOA_basics}

We now review the main aspects of QAOA, with a particular emphasis on the features most relevant to OpenQAOA. For the remainder of the article, we refer to the original implementation of QAOA in Ref.~\cite{Farhi2014} as \emph{standard} QAOA. Furthermore, given the extensions and modifications of the standard QAOA that have been developed, we will often refer to their collective as a family of QAOAs.

\subsection{Algorithmic components}

Given a cost Hamiltonian $H_c$, the objective of QAOA is to optimise the expectation value $\langle \Psi (\gamma,\beta) | H_c|\Psi (\gamma,\beta) \rangle$. In the standard QAOA, the variational ansatz state $| \Psi (\gamma,\beta)\rangle$ is defined as
\begin{equation}
    \label{eq:QAOA_standard_state}
    | \Psi (\gamma,\beta)\rangle = \prod_{k = 1}^{p} e^{-i\beta_{k} H_{\textrm{m}}} e^{-i \gamma_{k} H_{\textrm{c}}} | + \rangle^{\otimes n}.
\end{equation}

The circuit is initialised in the state $| + \rangle^{\otimes n}$, where $n$ is the number of qubits. Sets of unitary operators are then alternately applied to the state in a layer-wise manner. A layer in the QAOA circuit denotes the application of the complete set of unitaries, and the parameter $p$ corresponds to the number of layers in the circuit ansatz. Each layer is composed of the alternate evolution under unitaries generated by the cost Hamiltonian $H_{c}$ encoding the problem, and a mixer Hamiltonian $H_{\textrm{m}}$. These unitary operators are parametrised by the variational parameters $\{ \gamma_{k}\}$ and $\{ \beta_{k}\}$ respectively, with $k = 1, \ldots, p$. \newline 


\emph{Cost Operator}: The most commonly envisioned usage of QAOA is to tackle classical binary combinatorial optimisation problems. In this case, the elements of the cost Hamiltonian are derived directly from the cost function associated with the problem at hand. It is well known that these problems can be cast in the Ising form, i.e. as a set of interacting spin-$1/2$ particles or binary variables~\cite{AndrewIsing}. This means that the objective function can be encoded as the expectation value of a cost Hamiltonian $H_c$ of the form

\begin{equation}
    H_c = \sum_j h_j Z_j + \sum_{j,k} J_{jk} Z_jZ_k,
\label{eq:QUBO}
\end{equation}

where $Z_j$ denotes the Pauli-Z operator acting on the $j^{th}$ spin, $h_j$ is a bias term for single spins indicating their preference for one of the two possible operator eigenvalues, and $J_{jk}$ denotes the coupling strength between the spins $j$ and $k$. Equation~\ref{eq:QUBO} may also contain linear and quadratic terms that enforce constraints relevant to the problem. These constraints give a significant positive contribution to the energy if violated, but zero otherwise. When the entire problem can be formulated this way, we say we have a problem in quadratic unconstrained binary optimisation (QUBO) form \cite{Glover2019QuantumBA}. \newline


The unitary operator associated with the cost Hamiltonian can be generalised beyond its standard form -- i.e., as it appears in Eq.~\ref{eq:QAOA_standard_state} -- by increasing the number of variational parameters in the ansatz. The corresponding `cost operator' for the $L^{th}$ layer, $U_c^{(L)}$, then takes the form 

\begin{equation}
    U_c^{(L)} = \exp\Big(-i \Big[\sum_j \gamma_j^L h_j Z_j +  \sum_{j,k} \gamma_{jk}^L J_{jk} Z_j Z_k\Big]\Big). 
\end{equation}



Here, the coefficients $\{h_j\}$ and $\{J_{jk}\}$ are as defined in Eq.~\ref{eq:QUBO}, and $\{\gamma_j^L, \gamma_{jk}^L\}$ denote the variational circuit parameters for each term in the cost Hamiltonian for the $L^{th}$ layer. \newline

\emph{Mixer Operator}: The mixer operator is responsible for exploring the solution space, and creating interference between the computational basis states. In standard QAOA, the mixer operator is constructed from the Hamiltonian $H_m = -\sum_{i} X_i$, where $X_i$ is the Pauli-X operator acting on the $i^{th}$ qubit. As with the cost operator, this unitary `mixer operator' for the $L^{th}$ layer, $U_m^{(L)}$, can also be generalised from its standard form as

\begin{equation}
    U_m^{(L)} = \exp\Big(i \sum_{j}\beta_j^L X_j\Big),
\end{equation}

where $\{\beta_j^L\}$ are the variational parameters associated with each term in the mixer Hamiltonian for the $L^{th}$ layer. Different applications may call for alternative mixer operators, for example in the context of constrained optimisation, where it is desirable to explore a limited subspace of the full Hilbert space~\cite{NASA_XY, zhang2021qed}. \newline

\emph{Parametrisation}: It is possible to establish fixed relationships between different sets of the variational parameters, leading to many possible state ans\"{a}tze. We refer to a particular specification of these relationships as a `parametrisation'. For example, in the standard QAOA, we fix all parameters associated with the cost operator within a given layer to be equal, and similarly for the mixer operator. That is, we set $\gamma_j^L = \gamma_k^L = \gamma_{jk}^L \equiv \gamma^L$, and $\beta_j^L \equiv \beta^L$; we refer to this as the `standard parametrisation'. While this parameterisation is prevalent in much of the literature, alternative approaches have been proposed, and will be discussed further in Section~\ref{sec:The_OpenQAOA_Ansatz}.\newline

\emph{Parameter initialisation}: For a given parametrisation, one must select initial values for the corresponding parameters. We refer to this as `parameter initialisation', or simply `initialisation' when the context is clear. The initialisation strategy can strongly impact the performance of the algorithm. For example, assigning random initial values could lead to acute issues in the scalability of the optimisation procedure \cite{Google_Barren_Plateaus, cerezo_barren_plateaus}.\newline

\emph{Initial state}: Before applying the cost and mixer operations, the quantum circuit must be prepared in some initial state. In standard QAOA, one uses the equal superposition state $|+ \rangle^{\otimes n}$, however this may be undesirable for certain applications, e.g. constrained optimisation problems. 

\subsection{Computational procedure}

QAOA consists of a hybrid process where quantum and classical routines are performed in a loop. The algorithm itself can be divided into three main phases, summarised in Fig.~\ref{fig:qaoa_structure}.

\begin{itemize}
    \item \emph{Preparation phase} ---  A circuit ansatz is selected by specifying fixed problem inputs and variational parameters. The inputs include the cost Hamiltonian, the number of layers, the structure of the mixer layer, the parametrisation, and the initialisation strategy.
    
    \item \emph{Classical-quantum loop phase} --- The circuit is prepared and executed either on a QPU or on a simulator. After the execution, measurement data is obtained from a QPU or a simulator (the wavefunction is obtained if a wavefunction simulator is being used). The measurement outcomes are then used to compute the cost Hamiltonian expectation value corresponding to the current set of variational parameters. Based on this expectation value, the classical optimiser proposes an updated set of variational parameters to be used in the next iteration. The algorithm thus consists of a loop over the circuit's creation-execution, followed by optimisation of the variational parameters. This loop exits when the optimiser's stopping criterion is met, and the variational parameters are considered optimised. We refer to these final values of the parameters as the `optimised' or `optimal' parameters, and the corresponding output quantum state as the `optimised' or `optimal' state.
    
    \item \emph{Result phase} --- The third and final stage is the collection of the results and their parsing and presentation to the user. The raw output of QAOA is a distribution over all the computational basis states $\{ |z \rangle \}$, where $|z\rangle$ denotes an eigenstate of $H_{c}$, and the set of associated optimal parameters, which we denote $\{ \beta^*\}$ and $\{\gamma^*\}$. Therefore, finding a unique bitstring solution encoding the answer to the original QUBO problem involves using a post-selection scheme after completing the optimisation process. Common choices are either the most probable bitstring, or the one corresponding to the lowest expectation value measured, depending on the goals at hand.  
\end{itemize}

\begin{figure*}
\includegraphics[width=\textwidth]{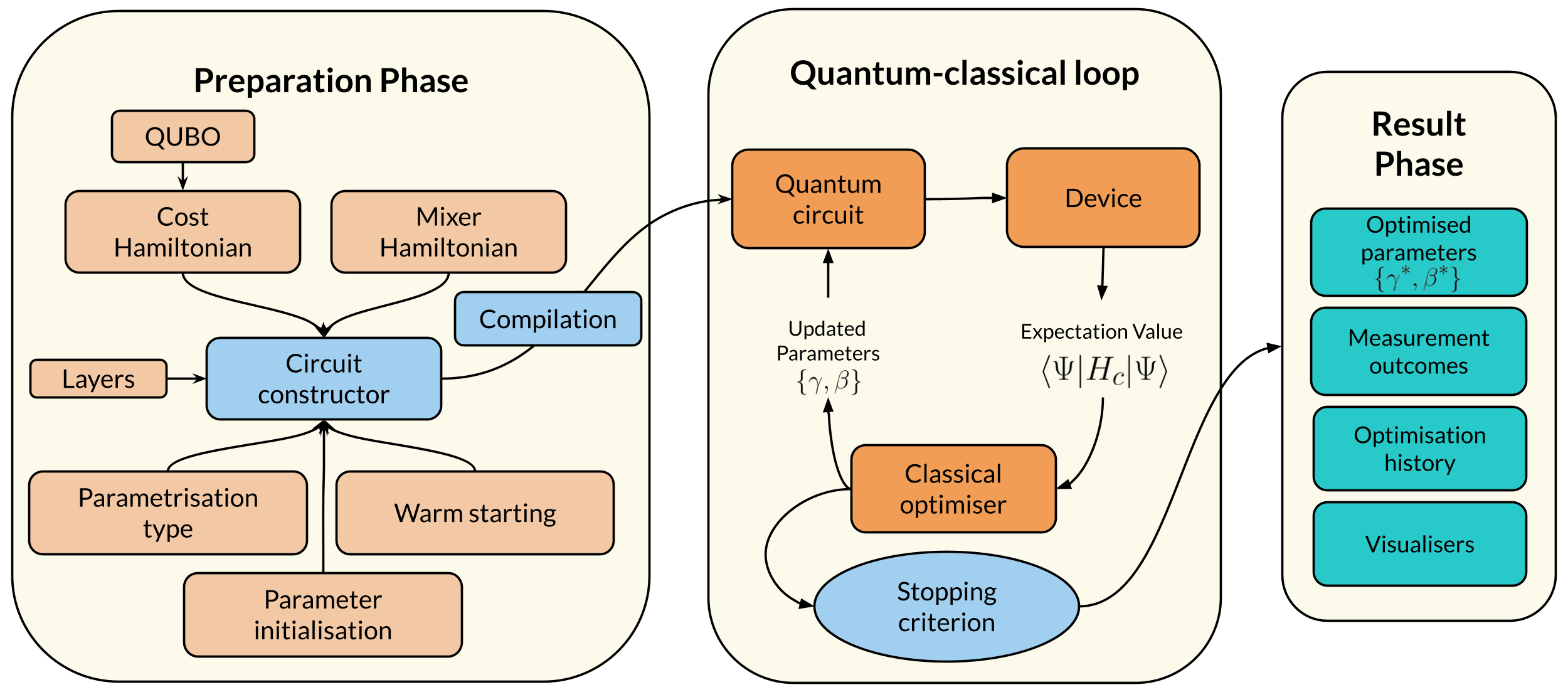}
\caption{The functional phases of OpenQAOA, and their relation to the different components of QAOA workflows. The preparation phase constructs the quantum circuit for the specified problem, with chosen input specifications. When a particular QPU backend is selected, the preparation phase includes a compilation step taking account of the device topology and native gate set. The quantum-classical loop phase is responsible for orchestrating the computation between the classical optimiser and the QPU or simulator, and terminates when a stopping criterion is reached. Following termination, the result phase provides a structure to access the problem solution and intermediate data associated with the optimisation.
}
\label{fig:qaoa_structure}
\end{figure*}

Several modifications, enhancements, and extensions of the standard QAOA have been proposed \cite{Hadfield2019AlternatingOperatorAnsatz, Zhu2022adaptive, Bravyi2020}. Intuitively, these modifications form a family of QAOAs, related in their structure through the layer-wise application of mixer and cost operations. Recursive QAOA (RQAOA) \cite{Bravyi2020}, which we introduce within the context of OpenQAOA in Section~\ref{sec:rqaoa}, is a prominent member of this family.


\section{OpenQAOA - API reference}

OpenQAOA provides APIs (application programming interfaces) between users and algorithms within the QAOA family. Interfaces are abstractions that help focus the user's attention on the QAOA, its implementation on QPUs, and its customisation by removing the need to write repetitive code dependencies to implement the algorithm explicitly. OpenQAOA interfaces are not constructed from a new circuit library, but are a programmable, abstract and device-independent description of QAOAs. \newline

In OpenQAOA, there are two types of interfaces: workflows and factory mode. Workflows have been designed to be the simplest way to execute an end-to-end QAOA. They consist of a series of logical steps enabling the user to select the type of QAOA and its properties, such as the parametrisation, compilation steps, or the choice of the classical optimiser. The factory mode trades simplicity for deeper manipulation of the objects used to build the workflows. Therefore, it is a way for users to either develop their own workflows or extract information that may not be directly exposed through the regular workflows. 



\subsection{QAOA workflows}

OpenQAOA comes with two pre-defined workflows: QAOA and RQAOA (see Section~\ref{sec:rqaoa}). Support for the Alternating Operator Ansatz~\cite{Hadfield2019AlternatingOperatorAnsatz} and ADAPT-QAOA~\cite{Zhu2022adaptive} is planned for future releases. While workflows implement QAOAs with the hybrid structure highlighted in Section~\ref{sec:QAOA_basics}, the API structure is slightly different. Given a QUBO problem, the simplest executable workflow is the following: \newline

\begin{code}
\begin{minted}{python}
from openqaoa.workflows import QAOA
q = QAOA()
q.compile(qubo)
q.optimize()
\end{minted}
\vspace{0.1cm}
\caption{For a given QUBO problem, this is the simplest OpenQAOA workflow. All variational parameters and other circuit properties here assume their default values.}
\label{code:simplestWorkflow}
\end{code}

A QAOA object is instantiated via the \emph{initialisation} step \verb|q = QAOA()|. In the simplest case, all the required variational parameters and circuit properties are set to default values. In the following sections, we will show how to customise these parameters and properties within the OpenQAOA workflow and implement different flavours of QAOA. This process includes customising circuit properties, choosing particular backends and devices, and configuring classical optimisers. \newline

The \emph{compilation} step \verb|q.compile(qubo)| is where the internal abstract representation of the QAOA is generated. The QUBO problem \verb|qubo| used in the compilation step is prepared either through the OpenQAOA problem classes located in \verb|openqaoa.problems.problem|, or it can be directly specified in terms of Python primitives by the user. OpenQAOA natively supports a set of popular problem classes modelled as QUBO problems. Note that, until this stage, the description of the QAOA is device independent and solely defined in terms of OpenQAOA primitives. \newline

The final stage is the \emph{optimisation} step, where the backend-specific circuit is created and submitted to either a simulator or a QPU. Subsequently, the optimisation loop is performed, and when the desired exit criteria are satisfied, the results will be accessible as attributes \verb|q.result_information| of the \verb|QAOA| object. \newline

The initialisation step is where the parameters of the QAOA can be modified to suit the user's needs. Codeblock~\ref{code:full_workflow} illustrates an example of how a workflow can be customised, and Sections~\ref{sec:The_OpenQAOA_Ansatz}~---~\ref{sec:result} are dedicated to explaining each customisation step in detail.

\begin{code}
\begin{minted}{python}
# Instantiate the QAOA object
q = QAOA()

# Define the circut ansatz
q.set_circuit_properties(
    p=2, 
    param_type='standard', 
    init_type='rand', 
    mixer_hamiltonian='xy'
    )
					
# Create and set the device
local_qasm_device = create_device(
    location='local',
    name='qiskit.qasm_simulator'
    )
q.set_device(local_qasm_device)

# Set the backend properties
q.set_backend_properties(
    n_shots=2000, 
    cvar_alpha=0.9,
    init_hadamard=True
    )

# Set the classical optimiser properties
q.set_classical_optimizer(
    method='nelder-mead', 
    maxiter=500,
    cost_progress=True
    )
					   
q.compile(qubo)
q.optimize()
\end{minted}
\vspace{0.1cm}
\caption{A customised workflow where the user sets the circuit ansatz, the parameter of the optimisation loop, and the backend properties. This demonstrates how the user can customise the workflows extensively with minimal effort.}
\label{code:full_workflow}
\end{code}

\subsection{The OpenQAOA Ansatz}
\label{sec:The_OpenQAOA_Ansatz}

In OpenQAOA, the circuit ansatz can be modified easily through the workflow routines. This is done by using the \verb|set_circuit_properties| method.\newline

The user can specify the type of circuit parametrisation and its initialisation through the keywords \verb|param_type| and \verb|init_type|, and the number of layers is set through \verb|p|. OpenQAOA offers several classes of parametrisation. \newline

\begin{code}
\begin{minted}{python}
q.set_circuit_properties(
    p=2, 
    param_type='standard', 
    init_type='rand', 
    mixer_hamiltonian='xy'
    )
\end{minted}
\vspace{0.1cm}
\caption{The circuit properties dictate the number of layers in the QAOA, the parameter type and their initialisation strategy, and the type of desired mixer Hamiltonian.}
\label{code:circuit_setter}
\end{code}

\begin{itemize}
\item The default is the \emph{Standard} parametrisation. In a given layer, each qubit is assigned the same mixer rotation angle $\beta$, and each term in the cost Hamiltonian is assigned the same angle $\gamma$. 

\item The \emph{Fourier} parametrisation is defined through the discrete cosine and sine transforms of the $\gamma$ and $\beta$ parameters across several layers, as proposed in Ref.~\cite{Zhou2020}. 

\item The \emph{Annealing} parametrisation emulates a discretised adiabatic annealing schedule using a QAOA circuit. 

\item Some parametrisations can further be generalised by increasing the number of free variational parameters. The \emph{With-Bias} sub-class parametrises the linear terms independently from the quadratic terms in the Hamiltonian. 

\item The \emph{Extended} sub-class parametrises each term in the Hamiltonian individually. In a standard QAOA circuit, there are $2p$ variational parameters, where $p$ is the number of layers. In the extended parametrisation, the number of parameters equals the number of terms in the Hamiltonian of Eq.~\ref{eq:QUBO}. For fully-connected cost Hamiltonians, where all variables appear together in quadratic terms, the total number of parameters is $O(n^2)$, where $n$ is the number of variables. In general, the connectivity of the cost Hamiltonian for classical optimisation problems is defined by the pairs of $(i,j)$ in the second term of Eq.~\ref{eq:QUBO}.
\end{itemize}

More information on the construction of these parametrisations may be found in the OpenQAOA documentation \cite{OQDocs}. \newline

Besides setting the parametrisation class, initial values for the parameters must also be provided at the beginning of the optimisation loop. OpenQAOA allows for three initialisation strategies: \emph{random}, \emph{custom}, and \emph{linear ramp}. The linear ramp strategy draws inspiration from quantum annealing, and linearly increases the values of the $\gamma$ parameters across the QAOA layers, while linearly decreasing the $\beta$ parameters~\cite{EntropicaQAOA, sack2021quantum}.

\subsection{The OpenQAOA Devices}
\label{sec:devices}

OpenQAOA builds circuits compatible with many devices. The quantum computing cloud services supported are Amazon Braket, IBMQ, and Rigetti Computing's Quantum Cloud Services. With access to different cloud providers, users can perform QAOA on a variety of simulators and QPUs supported by these organisations. Local simulators are also available. The user can select the desired backend devices through a workflow as follows:

\begin{code}
\begin{minted}{python}
# IBMQ credentials
ibm_credentials = {
    "api_token": "<ENTER API TOKEN HERE>",
    "hub": "IBMQ HUB",
    "group": "IBMQ GROUP",
    "project": "IBMQ PROJECT"
}

# instantiate the device
ibmq_cloud = create_device(
    location='ibmq', 
    name='ibm_nairobi',
    **ibm_credentials)

# Set the device
q.set_device(ibmq_cloud)

# Warm start QAOA with an "initial\_circuit"
def initial_circuit(*args) -> qiskit.QuantumCircuit:
    ...
    
# Set the backend properties    
q.set_backend_properties(
    n_shots=2000, 
    cvar_alpha=1,
    init_hadamard=False,
    prepend_state=initial_circuit(*args)
    )
\end{minted}
\vspace{0.1cm}
\caption{Users can obtain quick access to cloud-based hardware by passing the corresponding credentials. The workflows currently support devices provided by IBM Quantum, Amazon Braket, and Rigetti QCS.}
\label{code:device_setter}
\end{code}

The OpenQAOA \emph{backend} is responsible for creating and executing the device-specific circuit, computing the expectation value, and interacting with the classical optimiser. The method \verb|set_backend_properties()| can then be used to set common backend properties, such as the number of shots or the CVaR $\alpha$-parameter \cite{Barkoutsos2020ImprovingVQ}. It also enables the preparation of an arbitrary initial state through the \verb|prepend_state| parameter.\newline

OpenQAOA provides access to simulators from popular open-source libraries such as Qiskit's QASM Simulator and PyQuil's Statevector Simulator, as well as a alternative option referred to as \emph{vectorized}. The \emph{vectorized} backend is a fast, QAOA-specific simulator developed by Entropica Labs, primarily based on the common Python library NumPy~\cite{harris2020numpy}. 

\subsection{The OpenQAOA optimisation loop}
\label{sec:optimisation_loop}

OpenQAOA automatically runs the classical-quantum optimisation loop by invoking the method \verb|q.optimize()| as shown in Codeblock~\ref{code:simplestWorkflow}. The choice of optimiser lies with the user, and OpenQAOA supports all optimisers found in SciPy's \verb|minimize| method~\cite{2020SciPy}. \newline

\begin{code}
\begin{minted}{python}
# classical optimiser properties
q.set_classical_optimizer(
    method='nelder-mead', 
    maxiter=500,
    optimization_progress=True, 
    cost_progress=True,
    parameter_log=True
    )
\end{minted}
\vspace{0.1cm}
\caption{The user can choose the optimiser and configure other properties of the quantum-classical optimisation loop. Additionally, this method can be used to log the optimisation results and history.}
\label{code:setter_optimiser}
\end{code}

OpenQAOA features native implementation of custom optimisation methods, including gradient descent, root mean squared propagation (RMSProp), Newton descent, quantum natural gradient descent \cite{stokes2020quantum}, and simultaneous perturbation stochastic approximation (SPSA) \cite{spall1998overview}. The codebase also provides both approximate and exact methods to implement derivative computations. These include finite difference, SPSA, and the recently developed parameter shift ~\cite{li2017hybrid,mitarai2018quantum,Wierichs2022GeneralPR} and stochastic parameter shift methods for quantum circuits~\cite{banchi2021measuring}. \newline

Finally, we emphasise that the optimiser has flags for tracking and logging the optimisation status, which we anticipate to be useful for research and debugging purposes. We discuss the content of these logs in the next section.

\subsection{The OpenQAOA result}
\label{sec:result}

In its most basic form, the result of a QAOA computation consists of a list of measurement outcomes (or the optimal statevector, if a statevector simulator was employed), and the corresponding optimal variational parameters and optimal cost. In addition to these, OpenQAOA also returns the $k$ bitstrings associated with the $k$ most probable states in the optimised distribution, and the unique bitstring corresponding to the lowest cost function value observed in the entire optimisation process.  \newline

\begin{figure}
\includegraphics[width=\textwidth/2]{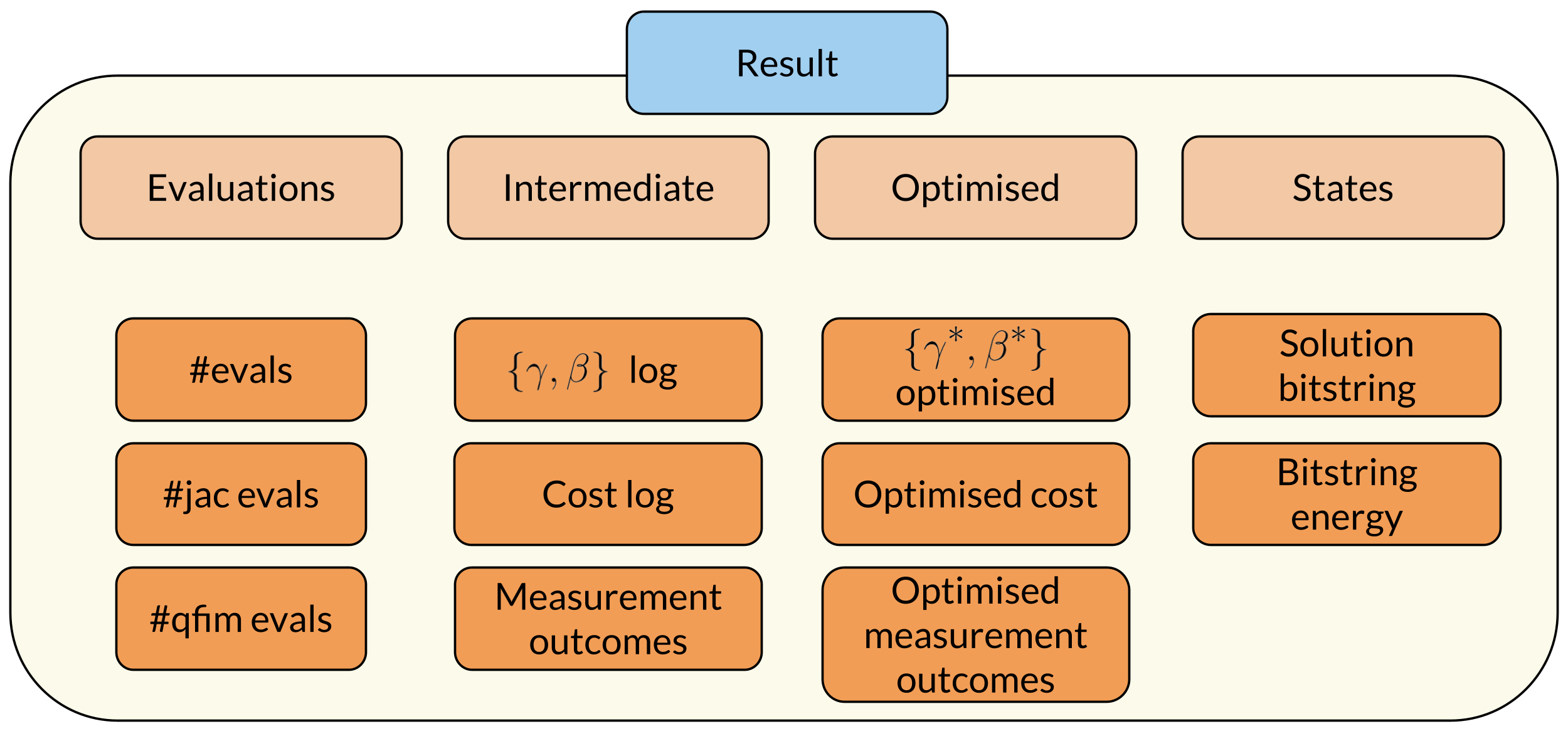}
\caption{To facilitate research, OpenQAOA provides a result object to store and access information about the optimisation run. At a high level, the results can be divided into number of evaluations, information on the intermediate states and outcomes, and on the final optimised state. The user can also access the probability distributions of the both optimised and intermediate circuits.}
\label{fig:oq_result_object}
\end{figure}

As an iterative algorithm, QAOA generates additional data during the optimisation loop, which can also be stored for later use. The OpenQAOA optimiser is thus equipped with logger methods to provide metrics from the classical optimisation process, such as the cost expectation value and the corresponding variational parameters for the $i^{th}$ iteration, and the measurement counts for the observed bitstrings. A summary of the attributes of the result object is given in Figure~\ref{fig:oq_result_object}. 

\subsection{The RQAOA workflow}
\label{sec:rqaoa}

RQAOA consists of running QAOA recursively and using the output statistics at each step to eliminate variables from the problem. When the reduced problem reaches a pre-defined cutoff size, it is solved exactly via classical methods. The final answer is then reconstructed from the solution obtained from this classical computation by re-inserting the eliminated variables in the appropriate order. Notably, the final output of RQAOA is a unique bitstring.\newline

In OpenQAOA, RQAOA is implemented through a workflow built on top of the QAOA workflow. In its simplest code formulation, it reads as follows:

\begin{code}
\begin{minted}{python}
# Define the QAOA properties
q = QAOA()

# Set type to custom, set steps to 1 and use the default QAOA
r = RQAOA(qaoa=q, rqaoa_type='custom')

# Set parameters for RQAOA. Here we fix the steps to 1 (default), the final cutoff value to 3
r.set_rqaoa_parameters(steps=1, n_cutoff=3)

# Compile and optimise
r.compile(qubo)
r.optimize()
\end{minted}
\vspace{0.1cm}
\caption{To implement Recursive QAOA, one first creates an instance of the QAOA object containing all desired properties, and then passes it as argument to the RQAOA workflow. All RQAOA-specific parameters can be set in a similar fashion as for the QAOA object. RQAOA is then compiled and optimised.}
\label{code:setter_optimiser}
\end{code}

OpenQAOA incorporates the RQAOA initially formulated in Ref~\cite{Bravyi2020}, and two generalised versions, which enable multiple variable (qubit) eliminations during the recursive process. These strategies are called \emph{custom} and \emph{adaptive}~\cite{EzeInProgressPaper}. The custom strategy allows the user to define the number of eliminations performed at each step. The adaptive strategy adaptively selects how many qubits to eliminate at each step based on a statistical criterion, with the user specifying the maximum number of eliminations allowed per step. \newline

Note how the RQAOA workflow requires a \verb|QAOA| object to be instantiated: this can be customised in the same way as a regular QAOA workflow, as described in the preceding sections.

\subsection{Factory mode}
\label{sec:factory_mode}

OpenQAOA workflows help orchestrate the entire QAOA routine on the user's backend of choice. They automate sequential creation and initialisation of different components that make up a QAOA computation, based on the inputs provided by the user. These components include the problem statement, variational parameters, device-specific backends and authentication, and classical optimiser initialisation. \newline

However, OpenQAOA also gives the user the option of building the QAOA computation manually, by-passing the pre-defined workflows. We refer to this as \emph{factory mode}, and Codeblock~\ref{code:manual_mode} presents an example to demonstrate the required steps. Factory mode allows greater flexibility, with more room for experimentation for advanced users. Since one of the central goals of OpenQAOA is to enable research, we consider this a significant feature. We now briefly discuss a few instances wherein a user may wish to employ the factory mode.\newline

\textbf{Using a custom classical optimiser}: OpenQAOA can facilitate creating the variational parameters and an authenticated backend with the ability to initialise QAOA circuits. The custom classical optimiser can subsequently call the expectation method of the backend, bypassing the need for the optimisers supported by OpenQAOA.\newline

\textbf{Plotting cost landscapes}: OpenQAOA backends support computing the cost value with respect to a set of parameter values (under the specified circuit parametrisation) for a QAOA circuit. A user may choose to define a grid of values for the parameters -- for example, $\beta$ and $\gamma$ for a $p=1$ standard QAOA circuit -- and obtain a plot of the cost landscape over the specified region. \newline

\textbf{More workflows}: Accessing the components directly, a user can build more complex workflows, for instance, a layerwise QAOA approach to optimise parameters for a depth-$p$ circuit.\newline

\textbf{Obtain the QAOA circuit}: With a greater degree of control in the factory mode, users may also choose to export the QAOA circuit. Note, however, that the circuit will be defined in the language of the chosen device. For instance, a \verb|qiskit.QuantumCircuit| if the selected device was an IBMQ QPU or simulator. \newline

Once the optimisation loop has terminated, the result is available as \verb|optimizer_obj.qaoa_result|, following a similar structure as the results obtained through a workflow.




\begin{code}
\begin{minted}{python}
# generate the cost and mixer Hamiltonian
terms = [(1,2),(2,3),(0,3),(4,0),(1,),(3,)]
coeffs = [1,2,3,4,3,5]
n_qubits = 5

cost_hamil = Hamiltonian.classical_hamiltonian(terms,coeffs)
mixer_hamil = X_mixer_hamiltonian(n_qubits)

# prepare the circuit parameters
qaoa_circuit_params = QAOACircuitParams(cost_hamil,mixer_hamil,p=10)
params = create_qaoa_variational_params(qaoa_circuit_params, 
						params_type='fourier',
						init_type='ramp',
						q=1)
 
# call the device to use
device_qiskit = create_device(location = 'local', name = 'qiskit.qasm_simulator')

# initialise the backend with the device and circuit\_params
backend_qiskit_p1 = get_qaoa_backend(circuit_params_p1, device_qiskit, n_shots = 500)

# prepare the optimiser
optimizer_dict = {'method': 'cobyla', 'maxiter': 100}
optimizer_obj = ScipyOptimizer(backend_obj, params, optimizer_dict)

# run the optimisation loop
optimizer_obj()
\end{minted}
\vspace{0.1cm}
\caption{OpenQAOA workflows provide a simple way to run QAOA problems on both simulators and QPUs. Underneath, they compile the problem following steps in \emph{factory mode} as shown here. For some use cases, an advanced user may bypass the workflow abstraction layer to access the factory mode. Some examples are provided in the text.}
\label{code:manual_mode}
\end{code}

\section{Conclusions}

We have introduced OpenQAOA, an SDK to execute QAOA workflows on QPUs and simulators. OpenQAOA helps the user to focus more on optimisation problems and the algorithmic components behind the QAOA, rather than on the tedious and repetitive tasks of writing quantum circuits from scratch. Through its abstract representation of QAOA workflows, OpenQAOA allows for a great degree of customisability, and provides compilation tools to extract maximum hardware performance. \newline

There are many challenges in scaling quantum algorithms to hundreds of qubits. Building cross-platform solutions that abstract complex algorithms into simple, human-friendly interfaces, is of paramount importance to the development of quantum computing. Scaling QAOAs to hundreds of qubits while lowering the entry barrier to run quantum applications are thus important steps towards making quantum computers easy to use, accessible, and useful. OpenQAOA aims to achieve exactly that, providing an open-source, community-oriented library to bring value to existing quantum devices, supporting a code base designed to empower researchers.

\section{Acknowledgements}

We would like to thank our former team members at Entropica who contributed to previous iterations of what has ultimately become OpenQAOA. In particular, we acknowledge Jan Lukas Bosse’s central role in the design and construction of the original EntropicaQAOA package \cite{EntropicaQAOA}.


\bibliographystyle{IEEEtran}
\bibliography{bibliography}

\appendix


\end{document}